\input{aipcheck}

\documentclass[
    ,final            
  ]
  {aipproc}

\layoutstyle{8x11single}

\begin{document}

\title{Indication of Two Classes in the Swift Short Gamma-Ray 
Bursts from the XRT X-Ray Afterglow Light Curves}

\classification{98.70.Rz}
\keywords      {gamma ray: bursts}

\author{T. Sakamoto}{
  address={CRESST and NASA Goddard Space Flight Center, Greenbelt, MD 20771}, 
  altaddress={Joint Center for Astrophysics, University of Maryand, Baltimore County, Baltimore, MD 21250}
}

\author{N. Gehrels}{
  address={NASA Goddard Space Flight Center, Greenbelt, MD 20771}
}

\begin{abstract}
We present the discovery of two distinct classes in the {\it Swift} short 
duration gamma-ray bursts (S-GRBs) from the X-Ray Telescope (XRT) 
X-ray afterglow light 
curve.  We find that about 40\% of the Swift S-GRBs have an X-ray 
afterglow light curves which only lasts less than 10$^{4}$ seconds after 
the burst trigger (hereafter short-lived S-GRBs).  On the other hand, 
another 60\% of S-GRBs have a long lasting X-ray afterglow light curve 
which resembles the long duration gamma-ray bursts.  We also find 
that none of the short-lived S-GRBs shows the extended emission in the 
Burst Alert Telescope (BAT) energy range.  We compare the burst 
properties for both the 
prompt emission and the afterglow, and discuss the possibility 
of different progenitors for the Swift short GRBs.  
\end{abstract}

\maketitle


\section{Introduction}

\begin{figure}[t]
\includegraphics[height=6.5cm]{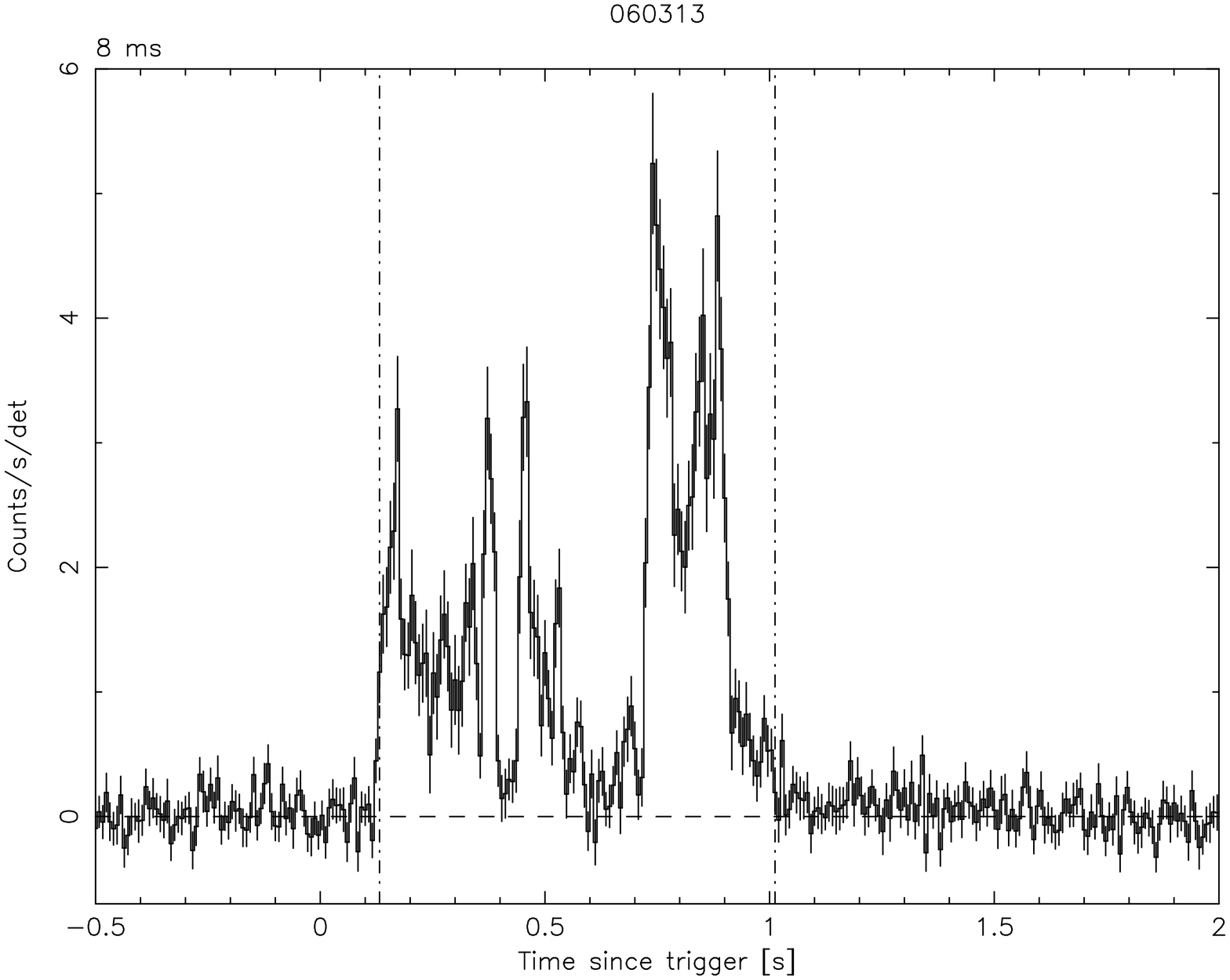}
\hspace{0.5cm}
\includegraphics[height=6.5cm]{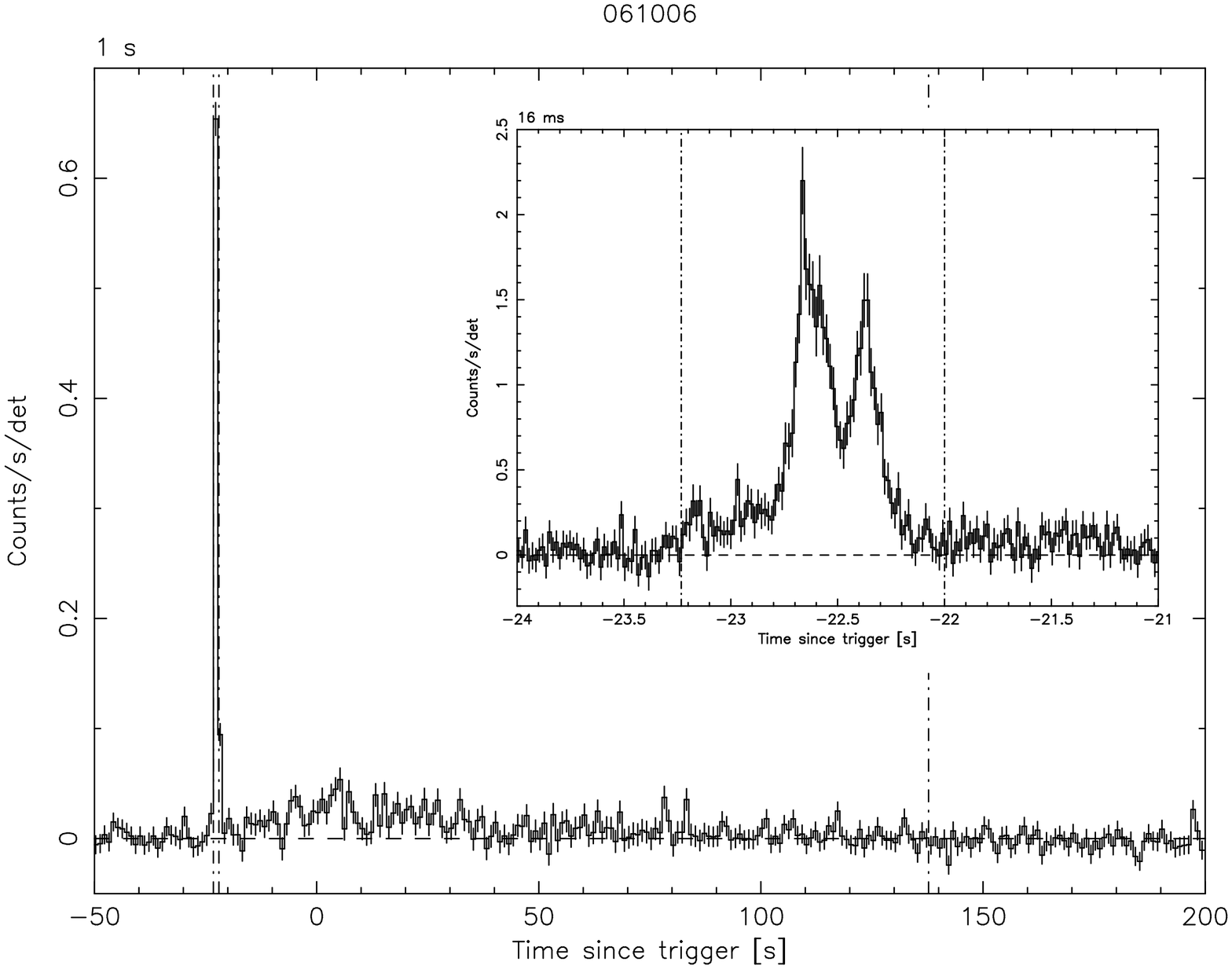}
\caption{\label{bat_lc} The BAT light curve of GRB 060313 as an example 
of S-GRBs without E.E. (left) and GRB 061006 as an example of S-GRBs with 
E.E. (right)  The insert of GRB 061006 is the BAT light curve around the 
initial short spike in 16 msec binning.}
\end{figure}

The distinct class in short duration GRBs (S-GRBs) has been claimed based 
on the prompt emission properties, namely a S-GRB with an extended 
emission (E.E.) \citep{norris2006} (see Figure \ref{bat_lc}).  The initial short spike 
of a S-GRB with an E.E. shows a negligible spectral lag which is one of the 
strong indications that the burst is indeed classified as a S-GRB 
\citep{norris2000}.  The E.E. emission tends to be softer than the initial 
short spike \citep[e.g.,][]{barthelmy2005,villasenor2005}.  

On the other hand, there is an indication for two different populations 
in S-GRBs based on the afterglow and the host galaxy properties.  For instance, 
the afterglow has been found only in X-rays for GRB 050509B.  The host galaxy 
of GRB 050509B is very likely to be an elliptical galaxy with no star formation 
\citep{gehrels2005}.  Whereas, in the case of GRB 051221A, the afterglow has 
been detected in all of frequencies (X-ray, optical and radio).  And its host 
galaxy is a star forming galaxy \citep{soderberg2005}.  

In this paper, we present the discovery of two distinct classes in 
the S-GRB X-ray afterglows observed by {\it Swift} 
X-ray Telescope (XRT).  We will discuss these two S-GRB classes comparing with 
their prompt emission properties, afterglows, and host galaxies.  

\section{Sample}
The samples of S-GRBs in our study include the bursts detected by {\it Swift} 
in 2005-2007.  We have 26 S-GRB samples in total;  19 of them have the X-ray 
afterglows; 9 of them have the optical afterglows; 17 of them have the redshift 
measurements (including 8 S-GRBs with host galaxy identifications only 
by the XRT position).  
Note that all of the redshifts of S-GRBs are from their host galaxy.  In this 
study, we will focus on 19 S-GRBs with X-ray afterglows (see Table \ref{sample_table}).  

The XRT light curves are collected from {\it Swift}/XRT GRB light curve 
repository \citep{evans2007}.  The analysis of the {\it Swift} Burst Alert 
Telescope (BAT) data has been performed using the standard BAT ftools.  

\begin{figure}[t]
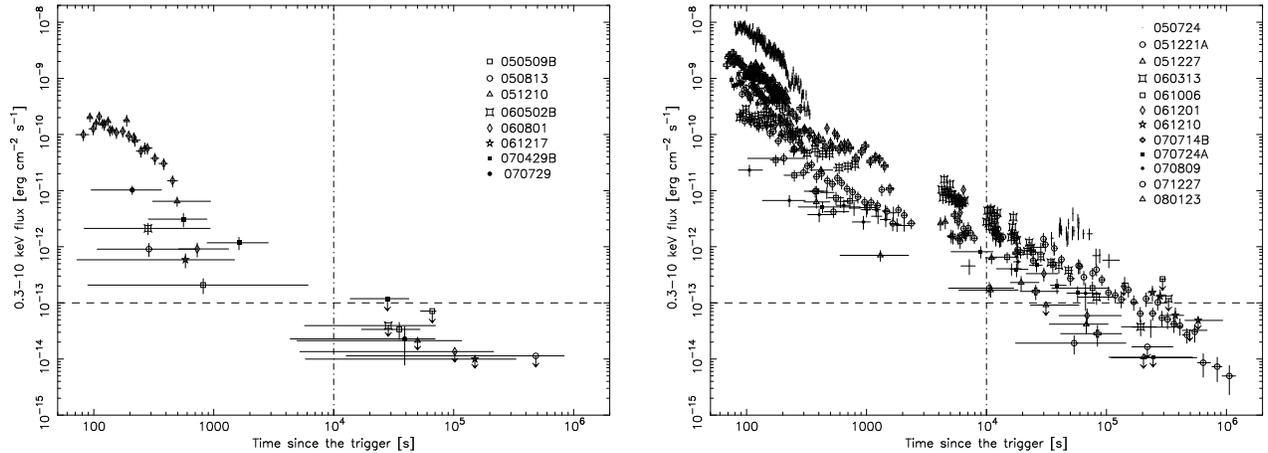

\includegraphics[height=8.0cm,angle=-90]{xrt_shortlive_bw.ps}
\hspace{0.5cm}
\includegraphics[height=8.0cm,angle=-90]{xrt_longlive_bw.ps}
\caption{\label{xrt_lc} The XRT X-ray afterglow light curves of the short-lived 
(SS; left) and the long-lived (LL; right) class.}
\end{figure}

\section{Short-lived and Long-lived X-ray Afterglow}

Figure \ref{xrt_lc} show the overlaid XRT X-ray light curve of S-GRBs.  
We notice that the X-ray light curves can be grouped into two 
classes.  The first class is ``short-lived (SL)'' X-ray afterglow; 
the X-ray flux is $<$ 10$^{-13}$ erg cm$^{-2}$ s$^{-1}$ at 10$^{4}$ sec 
after the trigger.  The second class is ``long-lived (LL)'' X-ray 
afterglow; the X-ray flux is $>$ 10$^{-13}$ erg cm$^{-2}$ s$^{-1}$ at 
10$^{4}$ sec after the trigger.  About 40\% of our S-GRB samples belongs 
to the SL X-ray afterglow class.  They have very faint X-ray afterglow 
lasting only a few hours after the trigger making them distinct 
compared to that of X-ray afterglow of L-GRBs.  On the other hand, the 
X-ray afterglow light curves of the LL class are similar to the L-GRBs.  
Table \ref{sample_table} shows whether 1) there is an E.E. detection in 
BAT (``E.E.'' column), 2) the optical afterglow (OA) has been detected 
(``OA'' column), and 3) the redshift has been measured (``redshift'' column).  
It is interesting to note 
that none of the SL S-GRBs have an E.E. and also an OA detection.  When we take 
into account that the redshifts of GRB 060502B and GRB 061217 are questionable, 
most of the S-GRB redshifts are from the LL S-GRB class.  

Figure \ref{dur_p64ms} shows the BAT peak count rate in 64 msec window versus 
the burst duration.  The peak count rate of all of the SL S-GRB class is small 
($<$ 0.5 counts s$^{-1}$ det$^{-1}$).  The LL S-GRB class has a mixture of a small 
and a large peak count rate.  We also notice that a peak count rate 
of a short spike varies even with or without an E.E.  
\citet{troja2008} found that all S-GRBs without E.E. lie far from the center of 
the host galaxy.  Figure \ref{dur_hgoffset} is the same figure in \citet{troja2008} 
by having different marks for the SL and the LL class.  Although it is not surprising 
because none of the SL class has an E.E. emission, most of the SL S-GRBs lie 
far from the center of the host galaxy.  

Based on the fact that the LL S-GRBs show 
1) similar X-ray afterglow light curve properties to L-GRBs, 2) are closer to 
the center of the host galaxy which is the characteristics of the L-GRB hosts, 
and 3) have detection of optical afterglows, the progenitor of S-GRBs with E.E. 
might be closer to L-GRBs.  

\section{Summary}

There is an indication of two classes in S-GRBs based on their X-ray afterglow 
properties.  There are the classes of S-GRBs with SL X-ray afterglows 
and LL X-ray afterglows.  The SL S-GRBs show 1) no E.E. and a small 
peak count rate in the prompt emission, 2) no optical afterglow, 3) a very small 
number of a redshift measurement, and 4) a larger offset from the center of the 
host galaxy.  Based on their characteristics, the afterglows and their host 
properties of the SL S-GRBs should provide a unique prove to understand the nature 
of S-GRBs.  

\begin{table}[htbp]
\begin{tabular}{lcccc|lcccc}
\hline
GRB     & Class & E.E. & OA & Redshift & GRB     & Class & E.E. & OA & Redshift\\\hline
050509B & SL    & N    & N  & Y        & 050724  & LL    & Y    & Y  & Y\\
050813  & SL    & N    & N  & N        & 051221A & LL    & N    & Y  & Y\\
051210  & SL    & N    & N  & N        & 051227  & LL    & Y    & Y  & N\\
060502B & SL    & N    & N  & Y(?)     & 060313  & LL    & N    & Y  & N\\
060801  & SL    & N    & N  & N        & 061006  & LL    & Y    & Y  & Y\\
061217  & SL    & N    & N  & Y(?)     & 061201  & LL    & N    & Y  & N\\
070429B & SL    & N    & N  & N        & 061210  & LL    & Y    & N  & N\\
070729  & SL    & N    & N  & N        & 070714B & LL    & Y    & Y  & Y\\
        &       &      &    &          & 070724A & LL    & N    & N  & Y\\
        &       &      &    &          & 070809  & LL    & N    & Y  & N\\
        &       &      &    &          & 071227  & LL    & Y    & Y  & Y\\\hline
\end{tabular}
\caption{\label{sample_table} The S-GRB sample in this work.  See text for the details.}
\end{table}

\begin{ltxfigure}
\begin{minipage}[t]{8cm}
\begin{center}
\includegraphics[height=8.0cm,clip,angle=-90]{dur_p64ms_bw.ps}
\caption[short caption for figure3]{\label{dur_p64ms} The BAT peak count rate in 64 msec 
window versus the burst duration.  The SL and LL are shown in triangles and circles, 
respectively.}
\end{center}
\end{minipage}
\hspace{0.5cm}
\begin{minipage}[t]{8cm}
\begin{center}
\includegraphics[height=8.0cm,clip,angle=-90]{dur_hgoffset_bw.ps}
\caption[short caption for figure4]{\label{dur_hgoffset} The projected host galaxy offset 
versus the burst duration at the GRB rest frame. The SL and LL are shown in triangles and 
circles, respectively.  The horizontal dashed line shows the averaged host galaxy offset 
of L-GRBs ($\sim$ 1.3 kpc).}
\end{center}
\end{minipage}
\end{ltxfigure}

\end{document}